\documentclass[aps,twocolumn]{revtex4}%
\usepackage{amsfonts}
\usepackage{amsmath}
\usepackage{amssymb}
\usepackage{graphicx}%
\begin{document}
\preprint{ }
\title{Muon-Spin Spectroscopy of the organometallic spin $1/2$ kagom\'{e}-lattice
compound Cu(1,3-benzenedicarboxylate)}
\author{Lital Marcipar, Oren Ofer, Amit Keren}
\affiliation{Physics Department, Technion, Israel Institue of Technology, Haifa 32000, Israel}
\author{Emily A. Nytko, Daniel G. Nocera, Young S. Lee, Joel S. Helton}
\affiliation{Department of Chemistry and Physics, Massachusetts Institue of Technology,
Cambridge, MA 02139 USA}
\author{Chris Baines}
\affiliation{Paul Scherrer Institute, CH 5232 Villigen PSI, Switzerland}
\pacs{PACS number}

\begin{abstract}
Using muon spin resonance we examine the organometallic hybrid compound
Cu(1,3-benzenedicarboxylate) [Cu(1,3-bdc)], which has structurally perfect
spin 1/2 copper kagom\'{e} planes separated by pure organic linkers. This
compound has antiferromagnetic interactions with Curie-Weiss temperature of
$-33$~K. We found slowing down of spin fluctuations starting at $T=1.8$~K, and
that the state at $T\rightarrow0$ is quasi-static with no long-range order and
extremely slow spin fluctuations at a rate of $3.6$~$\mu$sec$^{-1}$. This
indicates that Cu(1,3-bdc) behaves as expected from a kagom\'{e} magnet and
could serve as a model kagom\'{e} compound.

\end{abstract}
\date[Date text]{date}
\maketitle

The experimental search for an ideal two dimensional, spin 1/2, kagom\'{e}
compound, which has no out-of-plane interactions and no impurities on the
kagom\'{e} plane, has powered tremendous experimental effort in recent years
\cite{Samples}. Yet, all compounds studied today have shortcomings. Recently,
a promising copper-based metal organic hybrid compound Cu(1,3-bdc) was
synthesized by Nytko \emph{et al.} \cite{Emiliy}. This compound has an ideal
kagom\'{e} lattice structure as indicated by X-ray, the spins are naturally
1/2, and there are no Zn ions or any other candidates to substitute the Cu on
the kagom\'{e} plane. The goal of this paper is to show that from a magnetic
point of view Cu(1,3-bdc) shows the signatures of the high degree of
frustration expected on the kagom\'{e} lattice. This is done by demonstrating
that the inter-plane interactions are small enough compared to the intra-plane
interactions that no long range order is found at temperatures well below the
interaction energy scale, and by characterizing the ground state properties.
The experimental tool is muon spin resonance ($\mu$SR). Our major finding is
that the state at the lowest temperature investigated is quasi-static with
extremely slow spin fluctuations. This type of behavior is similar to a huge
class of frustrated magnets \cite{KagomeNoDeep,Canonical}. Therefore,
Cu(1,3-bdc) could serve as a model spin 1/2 kagom\'{e} compound.

\begin{figure}
[h]
\begin{center}
\includegraphics[
natheight=2.355700in,
natwidth=3.000000in,
height=2.38in,
width=3.0234in
]%
{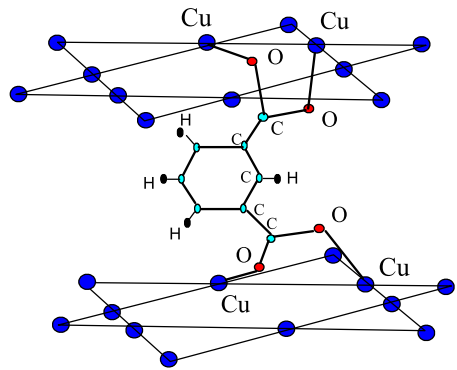}%
\caption{(Color online) The Cu(1,3-bdc) structure showing the kagom\'{e}
planes and the inter-plane and intra-plane superexchange path.}%
\label{compound}%
\end{center}
\end{figure}

Cu(1,3-bdc) is shorthand for Cu(1,3-benzenedicarboxylate). The kagom\'{e}
planes are separated by organic linkers, each linker being a benzene molecule
with two corners featuring a carboxylate ion instead of the standard H ion. If
one were to label the corners 1-6 consecutively, the two corners with the
carboxylate ions would be the 1st and 3rd. The Cu ions located on the
kagom\'{e} plane are linked by O-C-O ions, while inter-plane Cu ions are
linked by O-5C-O ions. The basic elements of Cu(1,3-bdc) are depicted in
Fig.~\ref{compound}. Magnetization measurements found antiferromagnetic
Curie-Weiss temperature of -33~K \cite{Emiliy}. Strong antiferromagnetic
exchange between Cu$^{2+}$ ions linked by a carboxylate molecule was also
found in the trinuclear compound Cu$_{3}$(O$_{2}$C$_{16}$H$_{23}%
$)\textperiodcentered1.2C$_{6}$H$_{12}$ \cite{StonePRB07}. Heat capacity shows
a peak at $T\simeq2$~K \cite{Emiliy}.

The powder we examined contains Cu(1,3-bdc) in the form of blue crystalline
plates. However, it is mixed with some green plates of copper-containing
ligand oxidation by-product C$_{32}$H$_{24}$Cu$_{6}$O$_{26}$, which cannot be
separated from the blue plates.\ To the naked eye it looks as if about 10\% of
the plates are green. However, as we demonstrate below, the $\mu$SR signal
from Cu(1,3-bdc) can be separated from the C$_{32}$H$_{24}$Cu$_{6}$O$_{26}$
signal. Our sample was pressed into a Cu holder for good thermal contact.%

%TCIMACRO{\FRAME{fhFU}{3.0234in}{2.38in}{0pt}{\Qcb{(Color online) The
%Cu(1,3-bdc) structure showing the kagom\'{e} planes and the inter-plane and
%intra-plane superexchange path.}}{\Qlb{compound}}{compound.eps}%
%width 3.0234in;  height 2.38in;  depth 0pt;  original-width 3in;
%original-height 2.3557in;  cropleft "0";  croptop "1";  cropright "1";
%cropbottom "0";  filename 'Compound.EPS';file-properties "XNPEU";}}}%
%BeginExpansion
%EndExpansion

Muon spin rotation and relaxation ($\mu$SR) measurements were performed at the
Paul Scherrer Institute, Switzerland (PSI) in the low temperature facility
spectrometer with a dilution refrigerator. The measurements were carried out
with the muon spin tilted at $45^{\circ}$ relative to the beam direction.
Positrons emitted from the muon decay were collected simultaneously in the
forward-backward (longitudinal) and the up-down (transverse) detectors with
respect to the beam direction. Transverse field (TF) measurements, where the
field is perpendicular to the muon spin direction, were taken at temperatures
ranging from 0.9~K to 6.0 K with a constant applied field of $H=1$~kOe.
Zero-field (ZF) measurements were taken in the longitudinal configuration at a
temperature ranging from 0.9~K to 2.8~K. The longitudinal-field (LF)
measurements, where the field is parallel to the muon spin direction, were
taken at several different fields between 50~Oe and 3.2~kOe with a constant
temperature of 0.9~K. We also performed a field calibration measurement using
a blank silver plate providing the muon rotation frequency $f_{s}=13.67$~MHz
at the applied TF of 1 kOe.%

%TCIMACRO{\FRAME{fhFU}{3.064in}{2.5339in}{0pt}{\Qcb{(Color online) FFT of the
%asymmetry data in a field of 1~kOe transverse to the initial muon spin
%dirrection. $f_{s}$ \ is the reference fequency in pure silver. Inset:
%transverse field asymmetry in the time domain and a rotating reference
%frame.}}{\Qlb{FFT}}{fft.eps}{\special{ language "Scientific Word";
%type "GRAPHIC";  display "USEDEF";  valid_file "F";  width 3.064in;
%height 2.5339in;  depth 0pt;  original-width 4.6095in;
%original-height 3.4091in;  cropleft "0";  croptop "1";  cropright "1";
%cropbottom "0";  filename 'fft.eps';file-properties "XNPEU";}}}%
%BeginExpansion
\begin{figure}
[h]
\begin{center}
\includegraphics[
height=2.5339in,
width=3.064in
]%
{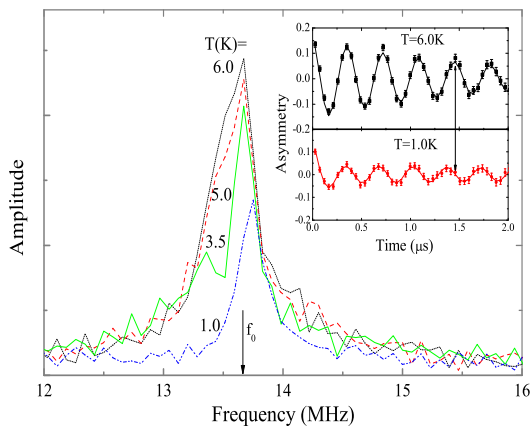}%
\caption{(Color online) FFT of the asymmetry data in a field of 1~kOe
transverse to the initial muon spin dirrection. $f_{s}$ \ is the reference
fequency in pure silver. Inset: transverse field asymmetry in the time domain
and a rotating reference frame.}%
\label{FFT}%
\end{center}
\end{figure}
%EndExpansion

In the inset of Fig. \ref{FFT} we depict by symbols the muon decay asymmetry
in a reference frame rotated at $H=200$~Oe less than the TF. In the main panel
of Fig.~\ref{FFT} we show the fast Fourier transform (FFT) of the TF data at
some selected temperatures. The FFT of the highest temperature, 6~K, shows a
wide asymmetric peak with extra weight towards low frequencies. At 3~K the
wide asymmetric peak separates into two different peaks shifting in opposite
directions. At even lower temperatures the low frequency peak vanishes. We
assign the latter peak to muons that stop in Cu(1,3-bdc) since such a wipe-out
of the signal is typical of slowing down of spin fluctuations, which in
Cu(1,3-bdc) is expected near $2$~K.

Despite the disappearance of the second peak in the frequency domain, its
contribution in the time domain is clear. The high frequency peak in the main
panel of Fig.~\ref{FFT} corresponds to the signal surviving for a long time in
both insets of Fig.~\ref{FFT}. The broad and disappearing peak in the main
panel corresponds to the fast decaying signal in the first 0.2~$\mu$sec seen
in the lower inset. The arrow in the inset demonstrates the frequency shift.
Consequently we fit the function
\begin{equation}
A_{TF}(t)=A_{1}e^{-(R_{1}t)^{2}/2}cos(\omega_{1}t+\varphi)+A_{2}e^{-R_{2}%
t}cos(\omega_{2}t+\varphi)+B_{g} \label{AsyTF}%
\end{equation}
to our data in the time domain globally, where the parameters $R_{1}$ and
$\omega_{1}$ are the relaxation and angular frequency of the by-product, and
$R_{2}$ and $\omega_{2}$ are the relaxation and angular frequency of the
kagom\'{e} part. The parameters $A_{1}=0.049(4),$ $A_{2}=0.125(3),$
$R_{1}=0.13(1)$~($\mu\sec^{-1}$)$,$ $\varphi$ and $B_{g}$ are shared in the
fit, while $R_{2},\omega_{1},\omega_{2}$ are free. The quality of the fit is
represented in the inset of Fig.~\ref{FFT} by the solid lines. The ratio of
$A_{1}$ to $A_{2}$ supports the assignment of the fast relaxing signal to Cu(1,3-bdc).%

%TCIMACRO{\FRAME{fhFU}{3.2474in}{2.3402in}{0pt}{\Qcb{(Color online) The muon
%shift from C$_{32}$H$_{24}$Cu$_{6}$O$_{26}$ $K_{1}$ and from Cu(1,3-bdc)
%$K_{2}$, and the relaxation rate from Cu(1,3-bdc) $R_{2}$ versus
%temperature.}}{\Qlb{KR}}{r2vsk.ps}{\special{ language "Scientific Word";
%type "GRAPHIC";  maintain-aspect-ratio TRUE;  display "USEDEF";
%valid_file "F";  width 3.2474in;  height 2.3402in;  depth 0pt;
%original-width 4.8594in;  original-height 3.4938in;  cropleft "0";
%filename 'R2VsK.ps';file-properties "XNPEU";}}}%
%BeginExpansion
\begin{figure}
[h]
\begin{center}
\includegraphics[
height=2.3402in,
width=3.2474in
]%
{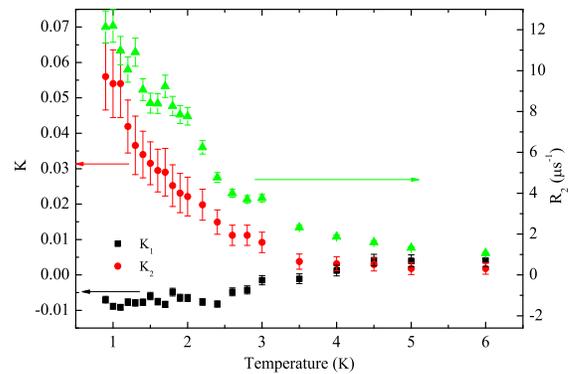}%
\caption{(Color online) The muon shift from C$_{32}$H$_{24}$Cu$_{6}$O$_{26}$
$K_{1}$ and from Cu(1,3-bdc) $K_{2}$, and the relaxation rate from Cu(1,3-bdc)
$R_{2}$ versus temperature.}%
\label{KR}%
\end{center}
\end{figure}
%EndExpansion

In Fig.~\ref{KR} we plot the shift, $K_{1,2}=(\omega_{s}-\omega_{1,2}%
)/\omega_{s}$, versus temperature, where $\omega_{s}=2\pi f_{s}$. As expected
$K_{2}$ increases with decreasing temperatures. The small decrease of $K_{1}$
is not expected and is not clear to us at the moment. The muon transverse
relaxation, $R_{2}$, is also presented in Fig. \ref{KR}. It has roughly the
same temperature behavior as the shift, $K_{2}$. However, at $T=$1.8~K $R_{2}$
seems to flatten out before increasing again around 1~K. This is somewhat surprising.

\begin{figure}
[h]
\begin{center}
\includegraphics[
trim=-0.020737in 0.031058in 0.020737in -0.031058in,
height=3.7887in,
width=2.9862in
]%
{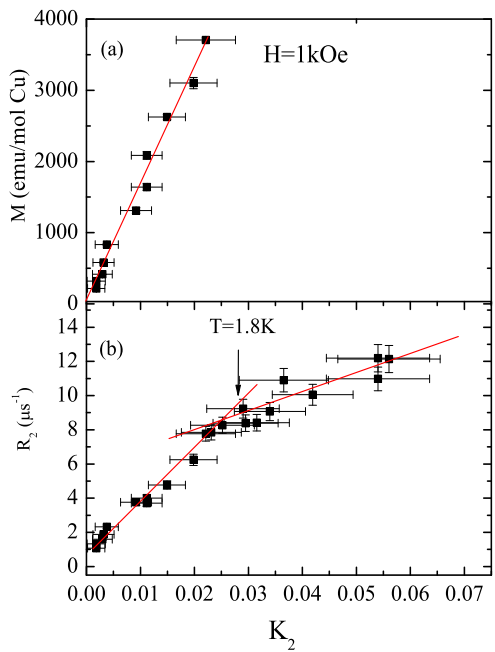}%
\caption{(Color online) (a) Magnetization versus the muon shift in
Cu(1,3-bdc). (b) The muon relaxation rate versus the muon shift in
Cu(1,3-bdc).}%
\label{ChiVsk}%
\end{center}
\end{figure}

The field at the muon site is given by $\mathbf{B}=\mathbf{H}-\sum_{k}%
\tilde{A}_{k}\left\langle \mathbf{S}_{k}\right\rangle $ where $\left\langle
\mathbf{S}_{k}\right\rangle $ is the thermal average of the spins neighboring
the muon, and $\tilde{A}_{k}$ is the hyperfine interaction with each
neighboring spin. Assuming a distribution of hyperfine fields in the $\hat{z}$
direction one can write $\tilde{A}_{k}$ as a sum of a mean value
$\overline{A_{k}}$ plus a fluctuating component $\delta A_{k}$. For the
distribution
\begin{equation}
\rho\left(  \delta A_{k}\right)  =\frac{1}{\pi}\frac{\sigma_{k}}{\left(
\delta A_{k}\right)  ^{2}+\sigma_{k}^{2}}%
\end{equation}
one finds that the shift is given by $K_{2}=\left\langle \mathbf{S}%
_{k}\right\rangle \sum_{k}\overline{A_{k}}/H$ and $R_{2}=\gamma_{\mu
}\left\langle \mathbf{S}_{k}\right\rangle \sum_{k}\sigma_{k}$
\cite{CarrettaHFM}. To test this derivation it is customary to plot the
macroscopic magnetization $M$ measured with a SQUID magnetometer versus
$K_{2}$. This is depicted in Fig.~\ref{ChiVsk}(a). The magnetization is also
measured at $1$~kOe. The plot indicates that in the temperature range where
both $M$ and $K_{2}$ are available they are proportional to each other.
Therefore, $\sum_{k}\overline{A_{k}}$ is temperature-independent.

If the $\sigma_{k}$ are also temperature-independent parameters we expect
$R_{2}\propto K_{2}$. A plot of $R_{2}$ versus $K_{2}$, shown in
Fig.~\ref{ChiVsk}(b), indicates that $R_{2}$ is not proportional to or even
does not depend linearly on $K_{2}$ and a kink is observed at $T_{0}=1.8$~K.
This result suggests a change in the hyperfine fields distribution at $T_{0}$.
An interesting possible explanation for such a change is a response of the
lattice to the magnetic interactions via a magnetoelastic coupling
\cite{KerenPRL01}. However, unlike a similar situation in a pyrochlore lattice
\cite{SagiPRL05}, it seems that here the lattice is becoming more ordered upon
cooling since the rate of growth of $R_{2}$ below $T_{0}$ is lower than at
higher temperatures.%

%TCIMACRO{\FRAME{fhFU}{2.9862in}{3.7887in}{0pt}{\Qcb{(Color online) (a)
%Magnetization versus the muon shift in Cu(1,3-bdc). (b) The muon relaxation
%rate versus the muon shift in Cu(1,3-bdc).}}{\Qlb{ChiVsk}}{chivsk.ps}%
%{\special{ language "Scientific Word";  type "GRAPHIC";  display "USEDEF";
%valid_file "F";  width 2.9862in;  height 3.7887in;  depth 0pt;
%original-width 4.4131in;  original-height 3.3814in;  cropleft "-0.004699";
%croptop "1.009185";  cropright "0.995301";  cropbottom "0.009185";
%filename 'ChiVsK.ps';file-properties "XNPEU";}}}%
%BeginExpansion
%EndExpansion

\begin{figure}
[h]
\begin{center}
\includegraphics[
height=3.2923in,
width=3.2517in
]%
{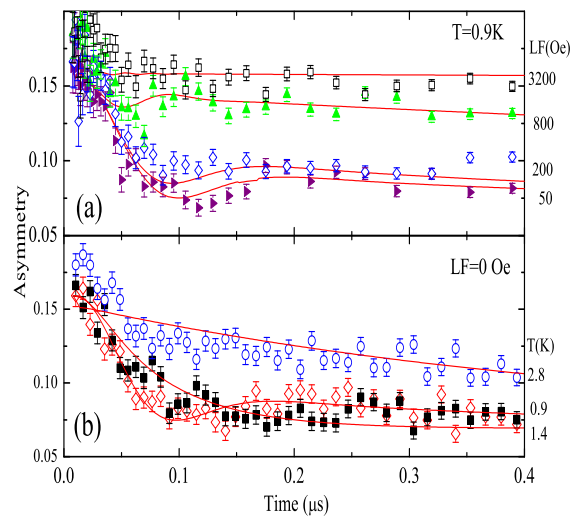}%
\caption{(Color online) The asymmetry at various longitudinal fields and
T=0.9~K (a). The asymmetry at zero field and various temperatures (b)}%
\label{LF}%
\end{center}
\end{figure}

The $\mu$SR LF data including ZF are presented in Fig.~\ref{LF}. The LF data
at the lowest temperature of $0.9$ K are depicted in panel (a). At this
temperature and a field of 50 Oe, the muon asymmetry shows a minimum at around
0.1~$\mu$sec. At longer times the asymmetry recovers. The origin of this dip
is the presence of a typical field scale around which the muon spin nearly
completes an oscillation. However, the field distribution is so wide that the
oscillation is damped quickly. The origin of the recovery is the fact that
some of the muons experience nearly static field in their initial field
direction during the entire measured time. These muons do not lose their
polarization while others do. When the external field increases, the dip moves
to earlier times (as the field scale increases) and the asymptotic value of
the asymmetry increases as well (as more muons do not relax).

The ZF data at three different temperature are shown in Fig.~\ref{LF}(b). As
the temperature decreases the relaxation rate increases due to the slowing
down of spin fluctuations, until at the lowest temperature the dip appears. We
saw no difference in the raw data between 1.0 and 0.9~K and therefore did not
cool any further.

These are unusual $\mu$SR data in a kagom\'{e} magnet, in the sense that the
spin fluctuations are slow enough compared to the internal field scale to
expose the static nature of the muon spin relaxation function, namely, the
dip, and to allow calibration of the internal field distribution. Other
kagom\'{e} magnets show the same general behavior but without this dip
\cite{KagomeNoDeep}. The data indicate the absence of long-range order and the
presence of quasi-static field fluctuation. If the ground state had long-range
order, the muon would have oscillated several times due to the internal
magnetic field. Similarly, if the ground state was dynamic we would \emph{not}
have seen a recovery of the muon polarization after a long time.%

%TCIMACRO{\FRAME{fhFU}{3.2517in}{3.2923in}{0pt}{\Qcb{(Color online) The
%asymmetry at various longitudinal fields and T=0.9~K (a). The asymmetry at
%zero field and various temperatures (b)}}{\Qlb{LF}}{lfzf.eps}%
%{\special{ language "Scientific Word";  type "GRAPHIC";  display "USEDEF";
%valid_file "F";  width 3.2517in;  height 3.2923in;  depth 0pt;
%original-width 4.9018in;  original-height 3.5483in;  cropleft "0";
%filename 'lfzf.eps';file-properties "XNPEU";}}}%
%BeginExpansion
%EndExpansion

To analyze this type of muon spin relaxation function, a theoretical
polarization function $P(\nu,\Delta,H,t)$ must be generated. It depends on the
random field distribution $\rho(\mathbf{B})$, the spin fluctuations rate $\nu$
defined by $\left\langle \mathbf{B}(t)\mathbf{B}(0)\right\rangle =\left\langle
B^{2}\right\rangle e^{-2\nu t}$ , where $B$ is the internal local field, and
the LF $H$. In ZF or small LF field, standard perturbation methods for
calculating relaxation functions do not apply and a special method for
calculating $P$ is required. This function is produced in two steps. In the
first step the static muon polarization is generated using the double
projection expression%
\begin{equation}
\overline{P}_{z}(0,\Delta,H,t)=\int\rho(\mathbf{B})\left[  \frac{B_{z}^{2}%
}{\mathbf{B}^{2}}+\frac{B_{x}^{2}+B_{y}^{2}}{\mathbf{B}^{2}}\cos(\gamma_{\mu
}\left\vert \mathbf{B}\right\vert t)\right]  d^{3}\mathbf{B.} \label{StatInt}%
\end{equation}
We found that the Gaussian field distribution
\begin{equation}
\rho(\mathbf{B})=\frac{\gamma_{\mu}^{3}}{(2\pi)^{3/2}\Delta^{3}}\exp\left(
-\frac{\gamma_{\mu}^{2}[\mathbf{B-}H\mathbf{\hat{z}}]^{2}}{2\Delta^{2}%
}\right)  \label{Rho}%
\end{equation}
works best. This $\overline{P}_{z}(0,\Delta,H,t)$ is known as the Static
Gaussian Kubo-Toyabe LF relaxation function \cite{Hayano}.

In the second step the dynamic fluctuations are introduced. One method of
doing so is using the Voltera equation of the second kind \cite{Keren}%
\begin{align}
\overline{P}_{z}(\nu,H,\Delta,t)  &  =e^{-\nu t}\overline{P}_{z}%
(0,H,\Delta,t)+\nonumber\\
&  \nu\int_{0}^{t}dt^{\prime}\overline{P}_{z}(\nu,H,\Delta,t-t^{\prime
})e^{-\nu t^{\prime}}\overline{P}_{z}(0,H,\Delta,t^{\prime}). \label{Voltera}%
\end{align}
The function $\overline{P}_{z}(0,H,\Delta,t^{\prime})$ is taken from the first
step. The factor $e^{-\nu t}$ is the probability to have no field changes up
to time $t$. The factor $e^{-\nu t^{\prime}}\nu dt^{\prime}$ is the
probability density to experience a field change only between $t^{\prime}$ and
$t^{\prime}+dt^{\prime}$. The first term on the r.h.s is the polarization at
time $t$ due to muons that did not experience any field changes. The second
term on the r.h.s is the contribution from those muons that experienced their
first field change at time $t^{\prime}$. The factor $e^{-\nu t^{\prime}%
}\overline{P}_{z}(0,H,\Delta,t^{\prime})\nu dt^{\prime}$ is the amplitude for
the polarization function evolving from time $t^{\prime}$ to $t$, which can
include more field changes recursively. This equation can be solved
numerically \cite{NumericalRecipes} and $\overline{P}_{z}(\nu,H,\Delta,t)$ is
known as the Dynamic Gaussian Kubo-Toyabe LF relaxation function \cite{Hayano}.

The experimental asymmetry is fitted with $A_{LF}=A_{0}\overline{P}_{z}%
(\nu,H,\Delta,t)+B_{g}$. The relaxation from the second green phase is very
small and is absorbed in the background factor $B_{g}$. In the fit of the
field-dependence experiment at the lowest temperature, presented in
Fig.~\ref{LF}(a) by the solid lines, $\Delta,\nu,A_{0}$ and $B_{g}$ are shared
parameters. We found $\Delta=19.8(4)$~MHz and $\nu=3.6(2)$~$\mu\sec^{-1}$.
This indicates that the spins are not completely frozen even at the lowest temperature.

\begin{figure}
[h]
\begin{center}
\includegraphics[
natheight=3.451500in,
natwidth=4.663900in,
height=2.9646in,
width=3.5016in
]%
{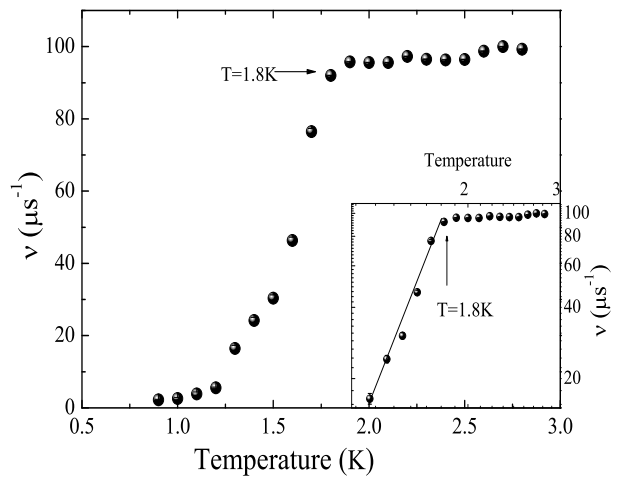}%
\caption{(Color online) The fluctuation rate $\nu$ versus temperature. Inset;
$\nu$ near $1.8$~K on a log-log scale. The error bars are smaller than the
symbol size.}%
\label{nu}%
\end{center}
\end{figure}

When analyzing the ZF data at a variety of temperatures, shown in
Fig.~\ref{LF}(b) by the solid lines, we permit only $\nu$ to vary. The fit is
good at the low temperatures but does not capture the 2.8~K data at early
times accurately. However, the discrepancy is not big enough to justify adding
more fit parameters. We plot the temperature dependence of the fluctuation
rate in Fig.~\ref{nu}. $\nu$ hardly changes while the temperature decreases
from $T=2.8~$K down to $T_{0}=1.8~$K. From $T_{0}$, $\nu$ decreases with
decreasing temperatures, but saturates below $1$~K. This type of behavior was
observed in a variety of frustrated kagom\'{e} (Ref.~\ \cite{KagomeNoDeep})
and pyrochlore (Ref. \cite{Canonical}) lattices. It is somewhat different from
classical numerical simulations where $\nu$ decreases with no saturation
\cite{kerenPRL94,RobertPRL08}. In fact, the numerical $\nu$ is a linear
function of the temperature over three orders of magnitude in $T$
\cite{RobertPRL08}.

The inset of Fig.~\ref{nu} shows $\nu$ as a function of temperature near
$T_{0}$ on a log-log scale where slowing down begins. Only near $T_{0}$ are
our data consistent with a linear relation%
\[
\nu-\nu_{\infty}=\nu_{0}(T-T_{0}),
\]
where $\nu_{\infty}$ is the high temperature fluctuation rate. The discrepancy
with the numerical work might be because $\mu$SR probes field correlations
involving several spins nearing the muon, while the simulations concentrate on
spin-spin auto correlations (with a decay $\Gamma_{a}$ compared here with
$\nu$). At our lowest temperature the rotations of ensemble of spins are
already coherent therefore field and spin correlations are not
identical.\ Another possibility is that the saturation of $\nu$ with
decreasing $T$ is a pure quantum effect not captured by the classical simulations.%

%TCIMACRO{\FRAME{fhFU}{3.5016in}{2.9646in}{0pt}{\Qcb{(Color online) The
%fluctuation rate $\nu$ versus temperature. Inset; $\nu$ near $1.8$~K on a
%log-log scale. The error bars are smaller than the symbol size.}}{\Qlb{nu}%
%}{nuvst.eps}{\special{ language "Scientific Word";  type "GRAPHIC";
%display "USEDEF";  valid_file "F";  width 3.5016in;  height 2.9646in;
%depth 0pt;  original-width 4.6639in;  original-height 3.4515in;
%cropleft "0";  croptop "1";  cropright "1";  cropbottom "0";
%filename 'NuVsT.EPS';file-properties "XNPEU";}}}%
%BeginExpansion
%EndExpansion

To summarize, we found that Cu(1,3-bdc) has a special temperature $T_{0}%
=1.8$~K. Upon cooling, the susceptibility, as measured by the $\mu$SR, grows
monotonically even past this temperature. The muon spin line-width also grows
but halts around this temperature. This might be explained by a subtle
structural transition, but low temperature structural data are required. At
$T_{0}$ the slowing down of spin fluctuations begins, but the spins remain
dynamic with no long range order. The rate of the spin fluctuations appears to
be linear near $T_{0}$, but becomes saturated at the lowest T. This general
behavior is similar to other kagom\'{e} compounds, though new features are
seen here. Therefore, considering its lattice, Cu(1,3-bdc) could serve as a
model compound for spin $1/2$ kagom\'{e} magnet.

We acknowledge financial support from the Israel U.S.A. Binational Science
Foundation, the European Science Foundation (ESF) for the `Highly Frustrated
Magnetism' activity, and the European Commission under the 6th Framework
Program through the Key Action: Strengthening the European Research Area,
Research Infrastructures. Contract n%
%TCIMACRO{\U{b0}}%
%BeginExpansion
${{}^\circ}$%
%EndExpansion
: RII3-CT-2004-506008.

\end{document}